\documentclass[prd,aps,superscriptaddress,twocolumn,floatfix,nofootinbib]{revtex4-1}
\pdfoutput=1

\usepackage{amsfonts}
\usepackage{amsmath}
\usepackage{amssymb}
\usepackage{bm}
\usepackage{dcolumn}
\usepackage{graphicx}   
\usepackage[latin1]{inputenc}
\usepackage{latexsym}
\usepackage{rotating}
\usepackage{hyperref}
\usepackage{graphicx}
\usepackage{color}

\usepackage{physics}
\usepackage{soul}
\usepackage{xcolor}
\usepackage{graphicx}% Include figure files
\usepackage{dcolumn}% Align table columns on decimal point
\usepackage{bm}% bold math
\usepackage{hyperref}% add hypertext capabilities
\usepackage[mathlines]{lineno}% Enable numbering of text and display math
\begin{document}

\title{Swampland Conjectures Constraints on Dark Energy from a Highly Curved Field Space}

\author{Guillaume Payeur}
\email{guillaume.payeur@mail.mcgill.ca}
\affiliation{Department of Physics, McGill University, Montr\'{e}al, QC, H3A 2T8, Canada}

\author{Evan McDonough}
\email{e.mcdonough@uwinnipeg.ca}
\affiliation{Department of Physics, University of Winnipeg,  Winnipeg MB, R3B 2E9, Canada}

\author{Robert Brandenberger}
\email{rhb@physics.mcgill.ca}
\affiliation{Department of Physics, McGill University, Montr\'{e}al, QC, H3A 2T8, Canada}

\date{\today}% It is always \today, today,
             %  but any date may be explicitly specified

\begin{abstract}

We study the interplay of the trans-Planckian censorship conjecture (TCC) and the swampland distance conjecture (SDC) in the context of multifield dark energy in a curved field space. In this scenario, the phase of accelerated expansion is realized as non-geodesic motion in a highly-curved field space, reminiscent of models developed in the context of inflation. The model features a stable attractor solution with near constant equation of state $w\simeq -1$, and predicts that the current era of accelerated expansion is eternal. The latter implies an eventual conflict with the TCC, which holds that the duration of any epoch of cosmic acceleration is bounded by the requirement that the large-scale observable universe is blind to Planck-scale early universe physics. This tension can be resolved by an interplay with the distance conjecture: for suitable parameter values, the apparent violation of the TCC occurs well after the fields have traversed a Planckian distance. The SDC then predicts a breakdown of the effective field theory (EFT) before the TCC can be violated. We derive the constraints on the model arising from the SDC+TCC and the de Sitter conjecture. We demonstrate that the model can be consistent with both swampland conjectures and observational data from Planck 2018 and the Dark Energy Spectroscopic Instrument.
\end{abstract}

%\keywords{Suggested keywords}%Use showkeys class option if keyword
                              %display desired
\maketitle
%\onecolumngrid

\section{Introduction}\label{sec:introduction}

The nature of dark energy is one of the key mysteries in fundamental physics.  While in principle dark energy could be a bare cosmological constant (see e.g. \cite{DE} for reviews of the dark energy mystery), this interpretation is facing both theoretical and recently also observational challenges. On the theoretical side, from an effective field theory point of view explaining dark energy by a cosmological constant would require extreme fine-tuning of the latter.  From the point of view of string theory,  there is increasing evidence, as will be reviewed in the next section, that no positive cosmological constant is possible.  From the observational point of view, there are some recent indications that the equation of state of dark energy is not that of a cosmological constant (see e.g. \cite{DESI, Reconstruct}), although the statistical evidence of this result is still weak. Thus, it is of interest to explore other possible explanations for dark energy.

Quintessence \cite{Peebles, Wetterich} has for a long time been an interesting possibility. The idea is that there is a new scalar field which is slowly rolling on a flat potential, resulting in an equation of state which can yield accelerated expansion of space.  A key question is the origin of such a scalar field.  Simple toy models of quintessence not based on any fundamental theory are not very satisfying.  Hence, it would be interesting to be able to embed a quintessence model in an ultraviolet complete theory such as superstring theory.  There have been a number of recent attempts to achieve this involving canonically normalized scalar fields arising from particular compactifications of string theory (see e.g. \cite{Cicoli, Heliudson}, and see \cite{cicoli_string_2023} for a review). However, although it is possible to obtain the correct energy scale for quintessence, the scalar field potentials which arise are generically too steep to allow an equation of state which is sufficiently close to that of a cosmological constant. 

In the context of inflation (where the requirements on the flatness of the potential are much more stringent) it has been proposed that inflation could be obtained with steep potentials if a multi-field model with a nontrivial field space metric in the kinetic term in the action is assumed \cite{Brown}.  A similar idea has recently been explored in order to obtain dark energy in the context of string theory compactifications \cite{Cicoli_new_2020, Cicoli_out_2020, brinkmann_stringy_2022}, as well as Early Dark Energy \cite{Heliudson2}, and it was shown that an effective field theory with the required structure may be obtained from string theory.

The ``trans-Planckian censorship conjecture''(TCC) \cite{TCC} implies that an effective field theory in which the dark energy phase of expansion does not terminate is inconsistent with fundamental theory (see e.g. \cite{TCCrevs} for discussions of this point which are independent of string theory).  In the context of string theory, the ``swampland distance conjecture'' (SDC) \cite{SDC} implies that the field range where a particular effective field theory is valid is bounded from above.  In this paper, we will study the interplay of these two conjectures in the context of a particular class of two field models with kinetic mixing which admit dark energy attractor solutions. We will identify the parameter space of this class of models which is consistent with both the TCC and the SDC. We find that there is a large range of parameter values where attractor solutions with an equation of state allowed by current observational constraints are possible (although this range of parameters may be outside of the region consistent with specific string theory compactifications).

In the following, we work in the context of a spatially flat homogeneous and isotropic cosmology with scale factor $a(t)$.  The Hubble expansion rate is denoted by $H(t)$, where $t$ is cosmic time, and $H_0$ is its current value. The radiation temperature is $T(t)$.  We work in units where the speed of light, Planck's constant and Boltzmann's constant are all set to 1. We denote the reduced Planck mass by $M_{pl} = (8\pi G)^{1/2} = 2.435\times 10^{18}$ GeV.

\section{Swampland Conjectures}\label{sec:swampland}

Recent work is highlighting the fact that not every effective quantum field theory (EFT) is consistent with a UV embedding in quantum gravity. The ``Swampland program'' aims to determine the constraints on effective field theory (EFT) necessary for a consistent UV completion in quantum gravity. The constraints are formulated in the form of a set of conjectures known as the ``Swampland conjectures'' (see \cite{Vafa, Palti, Valenzuela} for reviews).  Beginning with the ``Weak Gravity'' \cite{WGC} and ``Swampland Distance''\cite{SDC} conjectures, swampland conjectures now number in the dozens, pertaining to diverse field theories including dark photons models \cite{Reece:2018zvv} and models with gravitinos \cite{Kolb:2021nob}. 

Of particular relevance to cosmology is the ``de Sitter conjecture'' (\cite{obied_sitter_2018}), which asserts that an EFT including a possibly multi-component scalar field $\phi$ with potential energy function $V(\phi)$ weakly coupled to Einstein gravity must satisfy
\begin{equation}
    |\nabla V|/V \geq \frac{c}{M_{pl}}, \label{eq:dsc}
\end{equation}
where  $\nabla$ indicates the gradient in field space, and $c$ is a $\mathcal{O}(1)$ constant. The conjecture forbids de Sitter vacua since it forbids having $|\nabla V| = 0$ at positive energies $V>0$ and it also severely constrains inflationary models based on slow scalar field evolution \cite{Stein-Vafa, Lavinia}.\footnote{A refinement of the conjecture \cite{Garg:2018reu,Ooguri:2018wrx} allows $V'=0$ if $V''<0$. The models we consider will have $V''>0$ at all times.}  

Also pertinent to cosmology is the ``swampland distance conjecture'' (SDC) \cite{SDC}, which states that for an EFT including scalars $\varphi^I$ with kinetic term
\begin{align}
    S \supset  \int d^4x \sqrt{-g}\Big(\frac{1}{2}G_{IJ}(\varphi)\partial_\mu\varphi^I\partial^\mu\varphi^J\Big),
\end{align}
there is, in any large field distance limit, an infinite tower of states with mass scale $m$ decreasing exponentially as
\begin{align}
    m \sim M_{pl}e^{-\alpha d/M_{pl}},
\end{align}
where $\alpha$ is a $\mathcal{O}(1)$ constant and $d$ is the geodesic field distance from the point in moduli space wherein the original EFT is defined.  This means that the cutoff associated with this tower of states decreases exponentially with $d$. A consequence of this is that the EFT is only valid for geodesic distances bounded by
\begin{align}
    d \lesssim \frac{M_{pl}}{\alpha}\log\frac{M_{pl}}{\Lambda},
\end{align}
where $\Lambda$ is the cutoff of the EFT. The EFT therefore has a proper field range given by
\begin{align}
    \Delta\varphi^I = \mathcal{O}(1)M_{pl}, \label{eq:sdc}
\end{align}
beyond which the EFT breaks down.

There is a rich interplay between the distance and de Sitter conjectures \cite{Stein-Vafa,Dasgupta:2018rtp}. For example, if a model of a massive scalar field with potential $V(\phi)=m^2 \phi^2$ with $m^2 \ll M_{pl}^2$ were to lead to a period of inflationary expansion,  the de Sitter conjecture would be violated. But in such a model inflation would only occur if super-Planckian field excursions are possible, and this would violate the SDC.  Thus, the apparent violation of the de Sitter conjecture only would arise for field values which lie outside the realm of validity of the EFT. Our work will proceed in a similar spirit.

Related but distinct from the swampland conjectures is the ``trans-Planckian censorship conjecture'' (TCC) \cite{TCC}. This states that in a universe undergoing accelerated expansion,  the wavelength of quantum field fluctuations which initially have a length smaller than the Planck length should not become larger than the Hubble radius at any time.  As a consequence, any meta-stable de Sitter point has a lifetime $T$ bounded from above by
\begin{align}
    T = \frac{1}{H}\log\frac{M_{pl}}{H},
\end{align}
where $H$ is the Hubble parameter. The TCC places strict constraints on the allowed number of e-folds $N$ of dark energy-driven accelerated expansion in our universe's future.\footnote{It also leads to a stringent upper bound on the energy scale of a period of inflation which is sufficiently long to explain the origin of structure in the Universe \cite{TCC2}. }

The strictest bound on $N$ that can be placed without reference to uncertain early universe physics is that Planckian modes at the latest possible start to the universe's radiation-dominated era should remain sub-Hubble in the future. For consistency with Big Bang nucleosynthesis, the beginning of the radiation-dominated era must occur at a temperature $T_r \gtrsim 4$MeV \cite{hannestad_lowest_2004}. Using the convention that the scale factor $a$ is 1 today, the value of the scale factor $a_r$ at the start of the radiation-dominated era is determined by the Friedmann equation,
\begin{align}
    %H^2 = 
    H_0^2(\Omega_\Lambda + \Omega_m a_r^{-3} + \Omega_r a_r^{-4}) = \frac{1}{3M_{pl}^2}\frac{\pi^2}{30}g_\ast T^4 _r\label{eq:H},
\end{align}
where $H_0$ is the Hubble parameter today, the $\Omega$'s are the density parameters of dark energy, matter and radiation today, and again the subscript $r$ denotes the start of the radiation-dominated era. $g_\ast$ counts the number of relativistic degrees of freedom and is given by $g_\ast = 10.75$ at $T= 4$ MeV. The solution to eq.~(\ref{eq:H}) is given by $a_r = 4.4\times 10^{-11}$.

%\EM{
To track the evolution of length scales in the dark energy epoch we define a time coordinate $\tilde{N}$,
\begin{equation}
    \tilde{N} \equiv \ln \frac{k}{k_0} 
    \label{eq:Ntilde1},
\end{equation}
where $k_0\equiv a_0 H_0$ is the comoving wavenumber of a fluctuation which re-enters the Hubble radius today, and $k$ is the comoving wave number of a mode which enters at a later time $t$, when $a(t)H(t)=k$. We may alternatively express the above as
\begin{equation}
    \tilde{N}= \ln \frac{aH}{a_0 H_0},
\end{equation}
This is similar but distinct from the number of e-folds of expansion in the future,
\begin{equation}
    N \equiv \ln \frac{a}{a_0},
\end{equation}
where we define e-folds with respect to the present time. For a constant effective equation of state parameter of the cosmic fluid $w$, the two are related by
\begin{align}
    \tilde{N} = N\Big(1-\frac{3(1+w)}{2}\Big). \label{eq:N}
\end{align}
Now consider a mode which has physical wavenumber at the Planck scale at the beginning of Standard Hot Big Bang cosmology (the onset of radiation domination). This mode  has $k/a_r=M_{pl}$ and hence $k=a_r M_{pl}$. The requirement that this mode never re-enter the Hubble radius provides an upper bound on $\tilde{N}$ given by
\begin{equation}
    \tilde{N} \leq \log(\frac{a_rM_{pl}}{H_0}) \approx 115, \label{eq:Ntilde}
\end{equation}
where we have substituted $k=a_r M_{pl}$ into eq.~\eqref{eq:Ntilde1}. This places a bound on the number of e-folds of dark energy-driven accelerated expansion allowed in the future before a conflict with the TCC is certain to arise.

\section{Rapid Turn Dark Energy}\label{sec:rapid}

Accelerated expansion can be achieved in a single-field context by taking a sufficiently flat scalar potential $V(\phi)$ for the field $\phi$, e.g., the  ``quintessence'' model \cite{Copeland} (see also \cite{tsujikawa_quintessence_2013} for a review). Indeed, for a single field described as a barotropic perfect fluid in the presence of non-relativistic matter, models with constant 
\begin{align}
    \lambda \equiv -M_{pl}\frac{V'(\phi)}{V(\phi)},
\end{align}
have an attractor solution with a corresponding field equation of state 
\begin{align}
    w=-1+\lambda^2/3.
\end{align}
A requirement for single-field quintessence to yield accelerated expansion is then $\lambda \lesssim 1$ \cite{Lavinia}.  As discussed in section \ref{sec:swampland}, models with such a flat scalar potential may not be consistent with a UV embedding in a quantum gravity theory, motivating the search for alternative models.

\subsection{Multi-field dynamics}

Some multi-field models, on the other hand, allow accelerated expansion solutions without the requirement of a flat scalar potential. One such class of models is rapid turn models (see \cite{bjorkmo_rapid-turn_2019} for a review) which includes hyperinflation (\cite{Brown},\cite{bjorkmo_hyperinflation_2019}), sidetracked inflation \cite{garcia-saenz_primordial_2018} and angular inflation \cite{christodoulidis_angular_2019} (see also \cite{Aragam_Multi_2020}). All of the above rely on the negative curvature of field space in sustaining the accelerated expansion. While mostly discussed in the context of inflation, they are equally well applicable to dark energy \cite{Akrami_Multi_2021, Anguelova_Dark_2022, Anguelova_Dynamics_2024, Eskilt_Cosmological_2022}.

 Here we study a realisation of hyperinflation given by the two-field action
\begin{align}
    S = \int d^4x \sqrt{-g}\Big(\frac{1}{2}G_{IJ}(\phi)\partial_\mu\varphi^I\partial^\mu\varphi^J-V(\phi)\Big) \label{eq:action-general},
\end{align}
where $I,J=1,2$, and $\varphi^1 \equiv \phi$ and $\varphi^2 \equiv \psi$ are the two fields in the model. The model is characterized by a field space metric $G_{IJ}$ and a potential $V$ that both depend only on the field $\phi$. The field space metric $G_{IJ}(\phi)$ is given by
\begin{align}
    G_{IJ}(\phi) = \begin{pmatrix}
        1 & 0\\
        0 & e^{-\phi/(M_{pl}L)}
    \end{pmatrix}, \label{eq:field-space-metric}
\end{align}
corresponding to a surface of constant negative scalar curvature, with Ricci scalar
\begin{equation}
   M_{pl}^2 R_{\rm field-space}= - \frac{1}{2L^2}. \label{eq:R}
\end{equation}The parameter $L$ then controls the magnitude of the curvature of field space.

The scalar potential $V(\phi)$ is given by
\begin{align}
    V(\phi) = V_0 e^{-\lambda\phi/M_{pl}}.
\end{align}
The steepness of the potential is controlled by $\lambda$. Namely,
\begin{align}
    |\nabla V|/V = |\sqrt{G_{IJ}\partial_IV\partial_JV}|/V = \lambda/M_{pl}.
\end{align}
And thus for $\lambda \gtrsim {\cal O}(1)$ the model is in compliance with the de Sitter conjecture, eq.~(\ref{eq:dsc}).\footnote{This model has also been studied in \cite{Cicoli_new_2020}.}

In a flat FLRW universe, the action in eq.~(\ref{eq:action-general}), for the homogeneous part of $\phi$ and $\psi$, reduces to
\begin{align}
    S = \int d^4x \hspace{0.1cm} a^3\Big(\frac{1}{2}\dot{\phi}^2 + \frac{1}{2}e^{-\phi/(M_{pl}L)}\dot{\psi}^2 - V_0e^{-\lambda\phi/M_{pl}}\Big). \label{eq:action}
\end{align}
The classical equations of motion for $\phi$ and $\psi$ are then
\begin{eqnarray}
    \ddot{\phi} &&= -3H\dot{\phi} -\frac{1}{2M_{pl}L}e^{-\phi/(M_{pl}L)}\dot{\psi}^2 + \frac{V_0\lambda}{M_{pl}} e^{-\lambda \phi/M_{pl}} \nonumber \\
    \ddot{\psi} && = -\dot{\psi}\Big(3H-\frac{\dot{\phi}}{M_{pl}L}\Big) \iff \frac{d}{dt}(J) = 0,
\end{eqnarray}
where $J \equiv a^3 e^{-\phi/(M_{pl}L)}\dot{\psi}$ is the angular momentum in field space. In the regime where $\lambda L \ll 1 $ and $L \ll \lambda$, the model admits an attractor solution, which we refer to as the hyperbolic attractor, given by
\begin{align}
    \dot{\phi} &= 6M_{pl}HL \label{eq:attractor1}\\
    \dot{\psi} &= e^{\phi/(2M_{pl}L)}\sqrt{2V_0\lambda L e^{-\lambda \phi/M_{pl}}-(6M_{pl}HL)^2}.  \nonumber 
\end{align}
Moreover, the value of $w$ on the hyperbolic attractor is given by
\begin{align}
    w = \frac{\lambda L - 1}{\lambda L + 1} \approx -1 + 2\lambda L, \label{eq:w}
\end{align}
where we have used that $\lambda L \ll 1$. This makes it evident that accelerated expansion with $w \approx -1$ is possible with $\lambda \gtrsim 1$, provided that $\lambda L \ll 1$. Simply put, this model allows for accelerated expansion on steep potentials. Intuitively, this is possible because the effect of the kinetic coupling between $\phi$ and $\psi$ is to transfer kinetic energy from $\phi$ to $\psi$, and the latter is much more efficient at dissipating it through Hubble friction as a result of the factor $e^{-\phi/(M_{pl}L)}$ in eq.~(\ref{eq:field-space-metric}). We give a derivation of the hyperbolic attractor solution and the associated value of $w$ in appendix \ref{app:attractor}.

We now describe the behaviour of the fields $\phi$ and $\psi$ in a cosmological setting where they act together as dark energy and are coupled gravitationally to the matter in our universe. We first note that this model does not solve the coincidence problem, so parameters and initial conditions must be chosen such that the density parameter of $\phi$ and $\psi$, $\Omega_{(\phi,\psi)}$, matches the observed dark energy density, namely that $\Omega_{(\phi,\psi)}\approx 0.68$ \cite{Planck} today.  

\begin{figure}[h!]
    \centering
\includegraphics[width=0.5\textwidth]{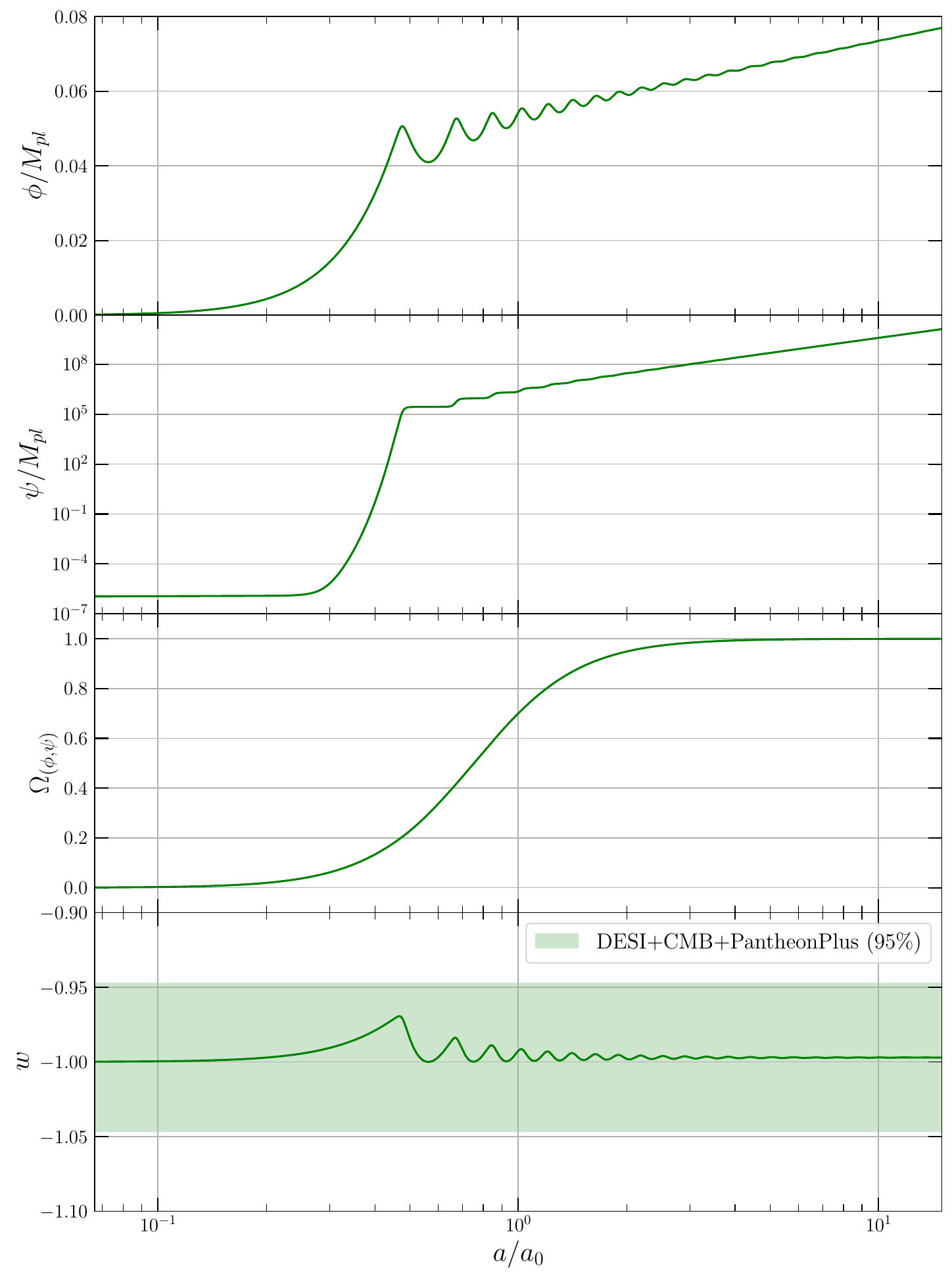}
    \caption{
 Field dynamics for $\lambda=1$ and $L=1/690$. $a$ is the scale factor chosen such that $a=a_0$ today. The fields are initially frozen through Hubble friction. In the vicinity of today, the density parameter $\Omega_{(\phi,\psi)}$ for $\phi$ and $\psi$ becomes order 1, whereupon $\phi$ and $\psi$ unfreeze and transition to the hyperbolic attractor solution towards which they converge at late times. Also shown is the equation of state parameter $w$ of the pair $(\phi,\psi)$.}
    \label{fig:field-dynamics}
\end{figure}

Under this assumption, the dynamics of $\phi$ and $\psi$ are as follows: In the early universe, the dynamics of both fields are dictated by Hubble friction, which freezes $\phi$ in place and rapidly redshifts any initial kinetic energy of $\psi$.
Around the present era, $\phi$ and $\psi$ unfreeze as a result of $\Omega_{(\phi,\psi)}$ becoming order 1, and transition to the hyperbolic attractor given by eqs.~(\ref{eq:attractor1}). In the far future, $\Omega_{(\phi,\psi)} \approx 1$, and the fields $\phi$ and $\psi$ continue to evolve on the hyperbolic attractor. There is a transition phase from the frozen state to the hyperbolic attractor which lasts a few e-folds of expansion,  during which the equation of state exhibits transient oscillations of amplitude that depend on model parameters and initial conditions. For illustration, figure \ref{fig:field-dynamics} shows a numerical evolution of $\phi$ and $\psi$ for model parameters $\lambda=1$ and $L=1/690$.

The equation of state parameter $w$ of the pair $(\phi,\psi)$ is approximately $-1$ at early times. At late times the fields evolve on the hyperbolic attractor, and consequently $w \approx -1+2\lambda L$. During the transition period, $w$ oscillates around the attractor value, with an amplitude that depends on parameters and initial conditions. This is shown in the bottom panel of figure \ref{fig:field-dynamics}.

In order to make contact with observational constraints on $w$, we make reference to the $w$CDM bounds on a constant $w$ from the combination DESI+CMB+PantheonPlus in \cite{desi_cosmological_2024}, namely
\begin{align}
    w = -0.997 \pm 0.025 \hspace{0.2cm}(68\%),
\end{align}
(see eq.~(5.3) from \cite{desi_cosmological_2024}). We approximate the 95\% confidence bound as 
\begin{align}
    w = -0.997 \pm 0.05 \hspace{0.2cm}(95\%). \label{eq:2sigma}
\end{align}
and ask that the hyperbolic attractor value of $w$ be inside the 95\% confidence interval of eq.~(\ref{eq:2sigma}), which results in the approximate constraint,
\begin{align}
    \lambda L \leq 0.0265 \label{eq:constraint-obs}.
\end{align}
from compliance with observational data. The example in figure \ref{fig:field-dynamics} is well within this observational bound. The amplitude of oscillations in $w$ in figure \ref{fig:field-dynamics} is small in comparison to the error bar in eq.~(\ref{eq:2sigma}), motivating {\it a posteori} the approximation of $w$ as a constant in our application of the observational bounds.

We note however, that the dynamics of the oscillations depend sensitively on the parameters $\lambda$, $L$ and the initial conditions (namely the state of the system prior to transitioning to the hyperbolic attractor), and the oscillations could potentially be of phenomenological interest. We leave a detailed analysis of the oscillatory phase to future work, and note they can be reduced in magnitude either by extending the model to generate the kinetic energy of $\psi$ dynamically in the late universe or by assuming a relatively large initial kinetic energy for $\psi$ (possibly the relic of an early kination phase, as conjectured to be a generic feature of string theory constructions \cite{Apers:2024ffe}).

\subsection{Appearance in string theory}

Effective actions of the same form as eq.~(\ref{eq:action}) can occur in string theory compactifications. A particular example is the case studied in \cite{saltman_scaling_2004}, where all closed string moduli are stabilized except for the overall Kahler modulus of a Calabi-Yau 3-fold. The four-dimensional effective action is $\mathcal{N}=1$ supergravity with Kahler potential
\begin{align}
    K = -3\ln(T+\bar{T}),
\end{align}
where $T = e^{\sqrt{2/3}\phi/M_{pl}} + i\sqrt{2/3}\psi /M_{pl}$ is the overall Kahler modulus. The kinetic part of the action is then
\begin{align}
    S \supset \int d^4x \sqrt{-g}\Big(M_{pl}^2K_{T\bar{T}}\partial_\mu T \partial ^\mu \bar{T}\Big),
\end{align}
which reduces to
\begin{align}
    S \supset \int d^4x \sqrt{-g} \Big( \frac{1}{2}(\partial_\mu \phi)^2 + \frac{1}{2}e^{-2\sqrt{\frac{2}{3}}\phi/M_{pl}}(\partial_\mu \psi)^2\Big).
\end{align}
In this scenario, the scalar potential is of the form
\begin{align}
    V = \frac{V_0}{(T+\bar{T})^3},
\end{align}
with $V_0>0$, which reduces to
\begin{align}
    V = \frac{V_0}{8}e^{-\sqrt{6}\phi/M_{pl}}.
\end{align}
The full action for $\phi$ and $\psi$ is then
\begin{align}
    S = \int d^4x \sqrt{-g} \Big(\frac{1}{2}(\partial_\mu \phi)^2 + \frac{1}{2}e^{-2\sqrt{\frac{2}{3}}\phi/M_{pl}}(\partial_\mu \psi)^2-V(\phi)\Big), \label{eq:action-string}
\end{align}
where
\begin{align}
    V(\phi) = \frac{V_0}{8}e^{-\sqrt{6}\phi/M_{pl}},
\end{align}
which, considering the homogeneous part of $\phi$ and $\psi$ in a FRLW universe, corresponds to the action of eq.~(\ref{eq:action}) for $\lambda=\sqrt{6}$ and $L=\sqrt{3/8}$. More generally, from a variety of string compactification scenarios where all moduli are stabilized save a single complex scalar, effective actions of the type
\begin{align}
    S = \int d^4x \sqrt{-g} \Big(\frac{1}{2}(\partial_\mu \phi)^2 + \frac{1}{2}e^{-2\sqrt{\frac{2}{\alpha}}\phi}(\partial_\mu \psi)^2-V(\phi) \Big),
\end{align}
where
\begin{align}
    V(\phi) = \frac{V_0}{2^{\alpha}}e^{-\sqrt{2\alpha}\phi},
\end{align}
are obtained. The action in eq.~(\ref{eq:action-string}) has $\alpha=3$, and in general, $\alpha$ is found to be a $\mathcal{O}(1)$ constant which, we emphasize, does not give rise to accelerated expansion satisfying observational constraints. We refer the reader to table 2 of \cite{brinkmann_stringy_2022} for a more exhaustive view of these scenarios\footnote{See also \cite{Aragam_Rapid_2022, Revello_Attractive_2024}.}. In studying instead the action of eq.~(\ref{eq:action}), we are converting the parameter $\alpha$ which is present in both the kinetic coupling and the scalar potential into two distinct free parameters $L$ and $\lambda$.

\section{Constraints from the Swampland}\label{sec:constraints}

We now find the constraints arising from conformity with the swampland conjectures reviewed in section \ref{sec:swampland} on the dark energy model described by eq.~(\ref{eq:action}). Specifically, we determine the region of the $(L,\lambda)$ parameter space which is in agreement with the dS conjecture and the TCC. We compare these constraints to those from observations and demonstrate that the model can be compatible with both swampland and observational constraints. The results are illustrated in figure \ref{fig:constraint}.
 
\subsection{Interplay between the TCC and SDC}

We begin with constraints from the trans-Planckian cosmic censorship. It was argued in section \ref{sec:swampland}
that the tightest constraint on the duration of accelerated expansion allowed before a conflict with the TCC is sure to arise is given by 
\begin{align}
    \tilde{N} \lesssim 115,
\end{align}
where $\tilde{N}$ represents the logarithmic interval in comoving wave number of the perturbation modes that cross the Hubble radius in the future during the dark energy era. That the above is an upper bound suggests a finite duration to the dark energy era. Meanwhile, the dark energy model, specified by the action eq.~(\ref{eq:action}), ostensibly predicts that accelerated expansion carries on indefinitely, as the system remains on the hyperbolic attractor solution with $w \approx -1 + 2\lambda L$ indefinitely, which would be in tension with the TCC. However, the situation changes once one also considers the Swampland Distance Conjecture.

The SDC predicts that the equations of motion have a proper field range beyond which they receive corrections. Analogous to the $\eta$ problem of inflationary cosmology \cite{Copeland:1994vg}, one expects these corrections to change the field dynamics, causing a premature end to the accelerated expansion. We consider the case where the order-1 constant in eq.~(\ref{eq:sdc}) is unity, and where the breakdown of EFT causes an end to the phase of accelerated expansion. Conformity with the TCC then amounts to the fields $\varphi^I$ moving a geodesic distance $d \geq M_{pl}$ prior to reaching $\tilde{N} = 115$.

Note that the TCC bound on the system is inherently path-dependent, namely, it depends on the evolution of the fields as encoded in the expansion history of the universe. In contrast, the SDC is strictly a statement of geodesic distance in field space. To study the interplay of the TCC and SDC we must first find a relation for the geodesic distance between any two points in field space, with metric given by eq.~(\ref{eq:field-space-metric}), such that the evolution of the fields can be translated into the evolution of the geodesic distance.

The geodesic equations are given by
\begin{align}
    \ddot{\phi}(s)+\frac{1}{2M_{pl}L}e^{-\phi(s)/(M_{pl}L)}\dot{\psi}^2(s)&=0\\
    \ddot{\psi}(s) -\frac{1}{M_{pl}L}\dot{\phi}(s)\dot{\psi}(s)&=0 \nonumber,
\end{align}
where $s$ parametrizes the geodesics. Their general solution, in the form $\phi(\psi)$, is given by
\begin{align}
    \phi(\psi) = M_{pl}L \log(C+D\psi-\frac{1}{4M_{pl}^2L^2}\psi^2) \label{eq:geo}.
\end{align}
where $C$ and $D$ are constants. The geodesic distance between two points in field space $(\phi=0,\psi=0)$\footnote{Without loss of generality we consider the initial field values at the current time to be $\phi = \psi = 0$. The choice $\phi = 0$ can always be achieved by rescaling $V_0$, and since the Lagrangian is symmetric under translations in $\psi$ direction, the choice $\psi = 0$ is justified.} and $(\phi=\phi_f,\psi=\psi_f)$ can then be computed via the integral
\begin{align} \label{eq:geod-exact}
    d &= \int_{0}^{\psi_f}d\psi\sqrt{\Big(\frac{d\phi(\psi)}{d\psi}\Big)^2+e^{-\phi(\psi)/(M_{pl}L)}}.
\end{align}
with $C$ and $D$ in eq.~\eqref{eq:geo} fixed such that $\phi(\psi)$ passes through both points. The exact result is given by eq.~\eqref{eq:geod-exact-app} in App.~\ref{app:geodesic}. This dramatically simplifies if we assume that the fields evolve from $(0,0)$ to $(\phi_f,\psi_f)$ on the hyperbolic attractor, and if we consider the limit $L\ll \phi_f/M_{pl}$. In that case, the geodesic distance can be approximated by
\begin{align}
    d \simeq \phi_f, \label{eq:geod-approx}
\end{align}
which, notably, is independent of $\psi_f$. More details regarding the derivation of this exact and approximate geodesic distance are given in appendix \ref{app:geodesic}.

\subsection{Constraints on the model}

\begin{figure}[h!]
    \centering
    \includegraphics[width=0.5\textwidth]{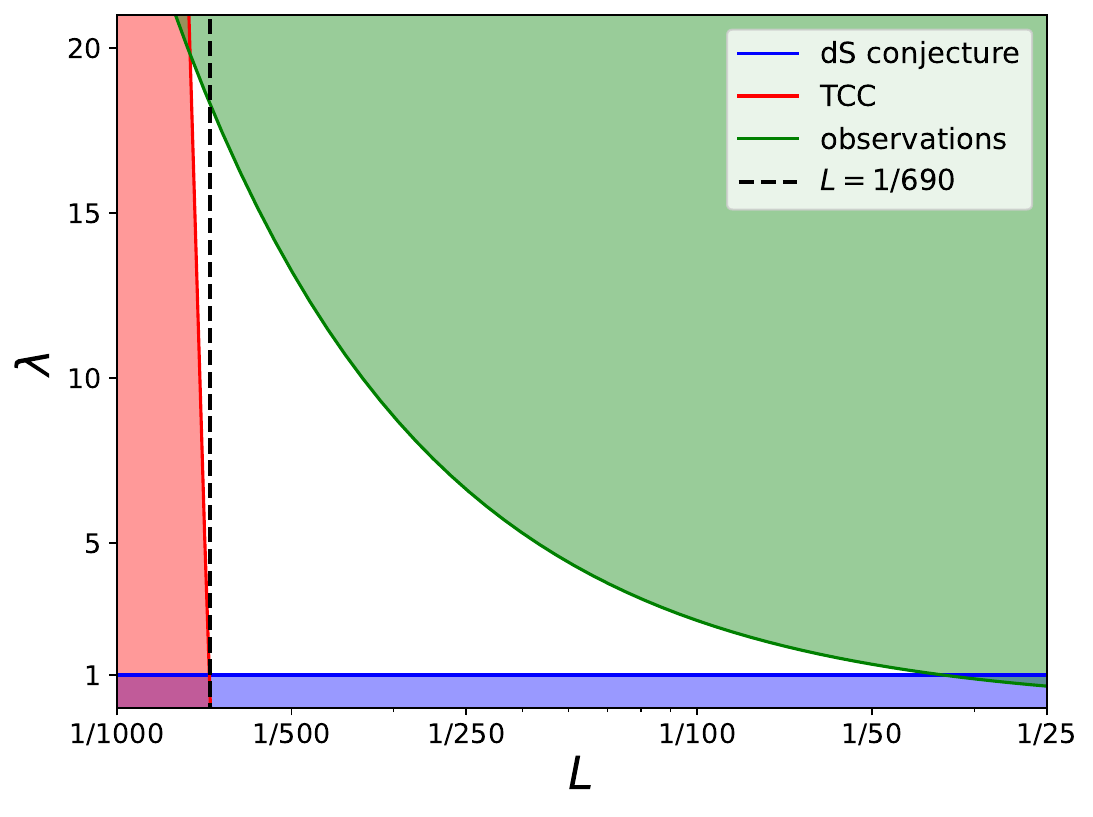}
    \caption{Constraints on $\lambda$ and $L$ arising from the TCC (red), the dS conjecture (blue), and observations (green). Shaded regions are excluded based on either the conjecture or the measurements. The white region corresponds to the region consistent with observational constraints and free of tension with the swampland conjectures.}
    \label{fig:constraint}
\end{figure}

Constraints on the model from both the swampland conjectures and from observation are shown in figure \ref{fig:constraint}. Each constraint is indicated by a solid curve and a shaded region indicating the region of parameter space that is excluded by said constraint. Here we study in detail the derivation of the constraints.

First, to determine the region of the $(L,\lambda)$ parameter space for which the fields $\varphi^I$ are displaced by a geodesic distance $d \geq M_{pl}$ before $\tilde{N} = 115$ is reached, we assume that the two-field system $\varphi^I = (\phi,\psi)$ lies on the hyperbolic attractor from today onwards. The validity of this assumption is justified in appendix \ref{app:simplifying}. By shifting $\phi$ and $\psi$ by a constant value, it is moreover possible to enforce that $\phi = \psi = 0$ at the time corresponding to today. This only requires a redefinition of the factor $V_0$ in eq.~(\ref{eq:action}). In that case, and given that $w=(\lambda L-1)/(\lambda L+1)$ is constant, we can solve the hyperbolic attractor solution $\dot{\phi}=6M_{pl}HL$ for $\phi$ as a function of $\tilde{N}$, which yields
\begin{align}
    \phi = \frac{12M_{pl}L}{-1-3w}\tilde{N}. \label{eq:phi-approx}
\end{align}
According to eq.~(\ref{eq:geod-approx}), the geodesic distance between the origin and the location of $\varphi^I$ at $\tilde{N}=115$ is then
\begin{align} \label{ineq}
    d = \phi|_{\tilde{N}=115} = \frac{12M_{pl}L}{-1-3w}115.
\end{align}
In order to be consistent with the TCC,  we want the value of $\phi$ when the SDC bound $d \leq 1$ is reached to occur before a conflict with the TCC arises. This yields the constraint
\begin{align}
    \frac{12M_{pl}L}{-1-3w}115 \geq M_{pl},
\end{align}
which, after substituting in $w=(\lambda L -1)/(\lambda L + 1)$, can be written as
\begin{align}
    690L(\lambda L +1) \geq -2\lambda L + 1.\label{eq:constraint}
\end{align}
Note that eq.~(\ref{ineq}) implies $L\ll \phi_f/M_{pl}$ which shows that the approximations we have made are self-consistent.

The TCC constraint eq.~(\ref{eq:constraint}) is a central result of this article and is shown by the red curve in figure \ref{fig:constraint}. Note that for $\lambda = \mathcal{O}(1)/M_{pl}$, the first terms on the left and right side of eq.~(\ref{eq:constraint}) are negligible, giving the approximate constraint
\begin{align}
    L \gtrsim \frac{1}{690}. \label{eq:constraint-approx}
\end{align}
This approximate constraint constitutes an upper bound on the magnitude of the curvature of field space.  Note that, if the $\mathcal{O}(1)$ constant in eq.~(\ref{eq:sdc}) is generalized to a constant $C$, the constraint reads $L \gtrsim C/690$.

We also impose the dS swampland conjecture, eq.~(\ref{eq:dsc}). We take the constant $c$ in eq.~(\ref{eq:dsc}) to be 1 and infer the constraint 
\begin{align}
    \lambda \geq 1,
\end{align}
which is shown by the blue curve in figure \ref{fig:constraint}. 

Finally, we also consider the approximate observational bounds on $\lambda$ and $L$ derived in eq.~(\ref{eq:constraint-obs}), shown by the green curve in figure \ref{fig:constraint}.

\section{Conclusions}\label{sec:conclusions}

In this work we have considered a two-field dark energy model with a curved field space metric. The model has two free parameters: the curvature of the field space metric and the slope of the scalar potential.  Treated as an effective field theory  (EFT), accelerated expansion would continue eternally and would eventually lead to a conflict with the trans-Planckian censorship conjecture (TCC).  However, the field range over which the EFT can be applied is bounded by the swampland distance conjecture (SDC).  Therefore, if the field values at which the TCC would begin to be violated were larger than the maximal field distance allowed by the SDC,  we would expect the EFT to break down, and hence a violation of the TCC could be avoided. We have identified the region in parameter space where both the de Sitter swampland conjecture can be satisfied and a conflict with the TCC avoided. This region of parameter space includes a subregion in which the observational constraints on the dark energy equation of state are obeyed.

It would be interesting to carefully study whether there are explicit string constructions which yield parameter values in the allowed region. Constructions which have been studied to date yield $\lambda L \sim 1$ \cite{gallego_anisotropic_2024,brinkmann_stringy_2022}, a value which is inconsistent with the observational constraints.

We note that there is a large mismatch between the geodesic distance which enters in the SDC and the field trajectory distance. It would be interesting to explore whether this observation has implications for previous studies of multi-field models of quintessence.

We also note that the negativity of the field space scalar curvature (see eq.~(\ref{eq:R})) causes an exponential growth of isocurvature perturbations \cite{Renaux-Petel_Geometrical_2016, Angelis_Adiabatic_2023, Fumagalli_Hyper_2019} (last reference is new). It was however found in \cite{Grocholski_Backreaction_2019} that in simple models of multi-field inflation in negatively curved field space, a kinematical backreaction effect shuts off the instability and preserves accelerated expansion. We leave a more detailed study of this for the model considered here for future work.

Another interesting direction for future work would be the application of the model presented here to {\it early} dark energy, building on string theory realizations of EDE presented in \cite{McDonough:2022pku,Cicoli:2023qri}. We leave this and other interesting possibilities to future work.

\section*{Acknowledgments}

The authors thank Stephon Alexander, Heliudson Bernardo, and Michele Cicoli, for helpful comments. The research at McGill is supported in part by funds from NSERC and from the Canada Research Chair program.  G.P. acknowledges support from the Fonds de Recherche du Qu\'ebec (FRQNT). E.M. is supported in part by a Discovery Grant from the Natural Sciences and Engineering Research Council of Canada, and by a New Investigator Operating Grant from Research Manitoba.

\appendix

\section{Derivation of hyperbolic attractor solution} \label{app:attractor}

Here we give a derivation of eqs.~(\ref{eq:attractor1}) which are an attractor solution to the background equations for motion for the action in eq.~(\ref{eq:action}) in the regime where $\lambda L \ll 1$ and $L \ll \lambda$. Recall that the EOM are given by
\begin{align}
    \ddot{\phi} = -3H\dot{\phi} -\frac{1}{2M_{pl}L}e^{-\phi/(M_{pl}L)}\dot{\psi}^2 + \frac{V_0\lambda}{M_{pl}} e^{-\lambda \phi/M_{pl}}\\
    \ddot{\psi} = -\dot{\psi}\Big(3H-\frac{\dot{\phi}}{M_{pl}L}\Big) \iff \frac{d}{dt}(J) = 0,
\end{align}
where $J=a^3 e^{-\phi/(M_{pl}L)}\dot{\psi}$. We search for a solution in which $\ddot{\phi}$ can be neglected. Then, the first EOM is
\begin{align}
    0 = -3H\dot{\phi} -\frac{1}{2M_{pl}L}e^{-\phi/(M_{pl}L)}\dot{\psi}^2 + \frac{V_0\lambda}{M_{pl}} e^{-\lambda \phi/M_{pl}},
\end{align}
which may be written in terms of $J$ as 
\begin{align}
    \frac{3M_{pl}H\dot{\phi}}{V_0\lambda e^{-\lambda\phi/M_{pl}}} - 1 &= -\frac{J^2}{2M_{pl}L}\frac{e^{\phi/M_{pl}L}}{a^6}\frac{M_{pl}}{V_0\lambda e^{-\lambda\phi/M_{pl}}}.
\end{align}
Now, make the assumption that $3H\dot{\phi} \ll (V_0\lambda/M_{pl}) e^{-\lambda\phi/M_{pl}}$, which is verified once the solution is found. Then,
\begin{align}
    &\frac{J^2}{2M_{pl3}L}\frac{e^{\phi/(M_{pl}L)}}{a^6}\frac{M_{pl}}{V_0\lambda e^{-\lambda\phi/M_{pl}}} \approx \text{const.}\\
    \implies &a^6 e^{-\phi/(M_{pl}L)} e^{-\lambda\phi/M_{pl}} \approx \text{const.}.
\end{align}
Taking the time derivative of both sides gives
\begin{align}
    \dot{\phi}\Big(\frac{1}{L}+\lambda\Big) &= 6M_{pl}H.
\end{align}
Since $1/L \gg \lambda$ this yields
\begin{align}
    \dot{\phi}=6M_{pl}HL,
\end{align}
which is eq.~(\ref{eq:attractor1}), the hyperbolic attractor solution for $\phi$. Using this, the first EOM becomes
\begin{align}
    0 &= -3H(6M_{pl}HL) - \frac{1}{2M_{pl}L}e^{-\phi/(M_{pl}L)}\dot{\psi}^2 + \frac{V_0\lambda}{M_{pl}} e^{-\lambda\phi/M_{pl}},
\end{align}
and solving for $\dot{\psi}$ gives
\begin{align}
    \dot{\psi} = e^{\phi/(2M_{pl}L)}\sqrt{2V_0\lambda L e^{-\lambda \phi/M_{pl}}-(6M_{pl}HL)^2},
\end{align}
which is the second line of eq.~(\ref{eq:attractor1}), the hyperbolic attractor solution for $\dot{\psi}$. We now verify that for the hyperbolic attractor solution, $3H\dot{\phi} \ll (V_0\lambda/M_{pl}) e^{-\lambda\phi/M_{pl}}$. Firstly, by substituting the hyperbolic attractor solution into the Friedmann equation, we have that
\begin{align}
    H^2 &= \frac{1}{3M_{pl}^2}\Big(\frac{1}{2}\dot{\phi}^2 + \frac{1}{2}e^{-\phi/(M_{pl}L)}\dot{\psi}^2 + V_0e^{-\lambda\phi/M_{pl}}\Big) \nonumber\\
    &= \frac{1}{3M_{pl}^2}\Big(V_0\lambda L e^{-\lambda\phi/M_{pl}}+V_0e^{-\lambda\phi/M_{pl}}\Big).
\end{align}
Using that $\lambda L \ll 1$, this gives
\begin{align}
    H^2 &\approx \frac{1}{3M_{pl}^2}V_0e^{-\lambda\phi/M_{pl}} = \frac{1}{3M_{pl}^2}V.
\end{align}
Using this, we obtain
\begin{align}
    \frac{3M_{pl}H\dot{\phi}}{V_0 \lambda e^{-\lambda\phi/M_{pl}}}  = \frac{6L}{\lambda} \ll 1,
\end{align}
where we used that $L \ll \lambda$. Finally, we can find the value of $w$ on the hyperbolic attractor solution. $w$ is given by
\begin{align}
    w &= \frac{\frac{1}{2}\dot{\phi}^2 + \frac{1}{2}e^{-\phi/(M_{pl}L)}\dot{\psi}^2 - V}{\frac{1}{2}\dot{\phi}^2 + \frac{1}{2}e^{-\phi/(M_{pl}L)}\dot{\psi}^2 + V},
\end{align}
which, upon substituting in the hyperbolic attractor solution for $\dot{\phi}$ and $\dot{\psi}$, yields
\begin{align}
    w = \frac{\lambda L-1}{\lambda L +1}.
\end{align}

\section{Geodesic distances derivations}\label{app:geodesic}

Here we give a thorough derivation of eq.~(\ref{eq:geod-exact}) for the geodesic distance between points $(\phi=0,\psi=0)$ and $(\phi=\phi_f,\psi=\psi_f)$ in a field space with metric given by eq.~(\ref{eq:field-space-metric}). We also derive the approximation given by eq.~(\ref{eq:geod-approx}).

The non-vanishing Christoffel symbols are
\begin{align}
    \Gamma^0_{\;\; 11} &= \frac{e^{-\phi/(M_{pl}L)}}{2M_{pl}L}\\
    \Gamma^1_{\;\; 01} &= \Gamma^1_{\;\; 10} = -\frac{1}{2M_{pl}L},
\end{align}
which leads to the geodesic equations
\begin{align}
    \ddot{\phi}(s)+\frac{1}{2M_{pl}L}e^{-\phi(s)/(M_{pl}L)}\dot{\psi}^2(s)&=0\\
    \ddot{\psi}(s) -\frac{1}{M_{pl}L}\dot{\phi}(s)\dot{\psi}(s)&=0.
\end{align}
which can be solved exactly as follows: By temporarily introducing the variable $u=\dot{\psi}(s)$, the second geodesic equation reduces to
\begin{align}
    u= \dot{\psi}(s) = C e^{\phi/(M_{pl}L)}.\label{eq:geo-psidot}
\end{align}
Substituting this into the first geodesic equation gives
\begin{align}
    \ddot{\phi}(s) + \frac{C^2}{2M_{pl}L}e^{\phi/(M_{pl}L)} = 0,
\end{align}
with solution
\begin{align}
    \phi(s) = M_{pl}L \log\Bigg(\frac{c_1\sech^2(\frac{\sqrt{c_1}}{2M_{pl}L}(s+c_2))}{C^2}\Bigg). \label{eq:geo-phi}
\end{align}
Substituting this back into eq.~(\ref{eq:geo-psidot}) yields
\begin{align}
    \dot{\psi}(s) = \frac{c_1\sech^2(\frac{\sqrt{c_1}}{2M_{pl}L}(s+c_2))}{C},
\end{align}
with solution
\begin{align}
    \psi(s) = \frac{2\sqrt{c_1}M_{pl}L}{C}\tanh(\frac{\sqrt{c_1}(s+c_2)}{2M_{pl}L})+D \label{eq:geo-psi}.
\end{align}
eqs.~(\ref{eq:geo-phi}) and (\ref{eq:geo-psi}) give the general solution in terms of arbitrary constants $C,D,c_1,c_2$. It can be organized into the form
\begin{align}
    \phi(\psi) = M_{pl} L \log(\frac{c_1}{C^2}-\frac{1}{4M_{pl}^2L^2}\psi^2+\frac{D}{2M_{pl}^2L^2}\psi - \frac{D^2}{4M_{pl}^2L^2}). \label{eq:geo-phiofpsi}
\end{align}
Here, one of the constants is redundant, and we can re-define them to write eq.~(\ref{eq:geo-phiofpsi}) as
\begin{align}
    \phi(\psi) = M_{pl} L \log(C+D\psi-\frac{1}{4M_{pl}^2L^2}\psi^2).
\end{align}
To find the geodesic passing between points $(\phi=0,\psi=0)$ and $(\phi=\phi_f,\psi=\psi_f)$, we set
\begin{align}
    &0 = M_{pl} L \log(C)\\
    \implies& C = 1,
\end{align}
and
\begin{align}
    &\phi_f = M_{pl} L \log(1+D\psi_f-\frac{1}{4M_{pl}^2L^2}\psi_f^2)\\
    \implies& D = \frac{e^{\phi_f/(M_{pl}L)}-1+\frac{1}{4M_{pl}^2L^2}\psi_f^2}{\psi_f}.
\end{align}
Therefore, the geodesic is given by
\begin{widetext}
\begin{align}
    \phi(\psi) = M_{pl}L \log(1+\frac{e^{\phi_f/(M_{pl}L)}-1+\frac{1}{4M_{pl}^2L^2}\psi_f^2}{\psi_f}\psi - \frac{1}{4M_{pl}^2L^2}\psi^2). \label{eq:geo-sol}
\end{align}
The geodesic distance between points $(\phi=0,\psi=0)$ and $(\phi=\phi_f,\psi=\psi_f)$ can then easily be computed, with the result
\begin{align}
    d &=  \int_{0}^{\psi_f}d\psi \sqrt{\Big(\frac{d\phi(\psi)}{d\psi}\Big)^2+e^{-\phi(\psi)/(M_{pl}L)}},
\end{align}
with $\phi(\psi)$ given by eq.~(\ref{eq:geo-sol}). This evaluates to
\begin{align}
    d=M_{pl} L \log(\frac{4(1+e^{\phi_f/(M_{pl}L)})M_{pl}^2L^2+\psi_f^2+\sqrt{16M_{pl}^4L^4(-1+e^{\phi_f/(M_{pl}L)})^2+8(1+e^{\phi_f/(M_{pl}L)})M_{pl}^2L^2\psi_f^2+\psi_f^4}}{4(1+e^{\phi_f/(M_{pl}L)})M_{pl}^2L^2+\psi_f^2-\sqrt{16M_{pl}^4L^4(-1+e^{\phi_f/(M_{pl}L)})^2+8(1+e^{\phi_f/(M_{pl}L)})M_{pl}^2L^2\psi_f^2+\psi_f^4}}) \label{eq:geod-exact-app}.
\end{align}
We can reduce this to a simpler approximate expression if we assume that the fields evolve from $(0,0)$ to $(\phi_f,\psi_f)$ on the hyperbolic attractor and that $L\ll \phi_f/M_{pl}$. As discussed in section   \ref{sec:constraints},  this is a good approximation for our considerations (see eq.~(\ref{ineq})). First, we can rewrite the denominator inside the log in eq.~(\ref{eq:geod-exact-app}) as
\begin{align}
    \text{den.} = 4(1+e^{\phi_f/(M_{pl}L)})M_{pl}^2L^2+\psi_f^2-\sqrt{16M_{pl}^4L^4(1-x+e^{\phi_f/(M_{pl}L)})^2+8(1+e^{\phi_f/(M_{pl}L)})M_{pl}^2L^2\psi_f^2+\psi_f^4}\Big|_{x=2}.
\end{align}
\end{widetext}
If evaluated at $x=0$, the denominator vanishes. With this in mind, using $L \ll \phi_f/M_{pl}$, we can Taylor expand to first order in $x$, around $x=0$, which reduces the denominator to the approximate expression
\begin{align}
    \text{den.}
    =\frac{32M_{pl}^4L^4e^{\phi_f/(M_{pl}L)}}{4M_{pl}^2L^2e^{\phi_f/(M_{pl}L)}+\psi_f^2}.
\end{align}
As for the numerator inside the log in eq.~(\ref{eq:geod-exact-app}), using $L \ll \phi_f/M_{pl}$, the expression approximates to
\begin{align}
    \text{num.} &= 2(4M_{pl}^2L^2 e^{\phi_f/(M_{pl}L)} + \psi_f^2).
\end{align}
We thus see from eq.~(\ref{eq:geod-exact-app}) that the geodesic distance is given approximately by
\begin{align}
    d &= M_{pl}L \log(\frac{(M_{pl}^2L^2e^{\phi_f/(M_{pl}L)}+\frac{1}{4}\psi_f^2)^2}{M_{pl}^4L^4e^{\phi_f/(M_{pl}L)}}) \nonumber\\
    &= M_{pl} L \log(e^{\phi_f/(M_{pl}L)}\Big(1+\frac{1}{4M_{pl}^2}\frac{\psi_f^2}{L^2e^{\phi_f/(M_{pl}L)}}\Big)^2).\label{eq:geod-intermediate}
\end{align}
To simplify this further, observe that the hyperbolic attractor solution yields
\begin{align}
    \dot{\psi}^2 &= 2V_0\lambda L e^{\phi(1/L-\lambda)/M_{pl}}-(6M_{pl}HL)^2e^{\phi/(M_{pl}L)}\nonumber\\
    &\approx 6H^2M_{pl}^2\lambda L e^{\phi/(M_{pl}L)}-(6M_{pl}HL)^2 e^{\phi/(M_{pl}L)}\nonumber\\
    &\approx e^{6Ht}(6H^2\lambda LM_{pl}^2),
\end{align}
where we used that $w\approx -1$ on the hyperbolic attractor, and that $\lambda L \ll L^2$. After integrating, this yields
\begin{align}
    \psi = \frac{\sqrt{6\lambda L}M_{pl}}{3}e^{3Ht},
\end{align}
and therefore,
\begin{align}
    \frac{\psi_f^2}{L^2 e^{\phi_f/(M_{pl}L)}}=\frac{2\lambda M_{pl}^2}{3L}.
\end{align}
The expression in eq.~(\ref{eq:geod-intermediate}) therefore reduces to
\begin{align}
    d &= M_{pl}L \log(e^{\phi_f/(M_{pl}L)}\Big(1+\frac{\lambda}{6L}\Big)^2)\nonumber\\
    &\approx M_{pl}L \log(e^{\phi_f/(M_{pl}L)}) + 2L\log(\frac{\lambda}{6L})\nonumber\\
    &= \phi_f + 2M_{pl}L\log(\frac{\lambda}{6L}). \label{eq:geod-intermediate2}
\end{align}
Finally, we show that $2M_{pl}L\log(\frac{\lambda}{6L}) \ll \phi_f$ allowing us to ignore the second term in eq.~(\ref{eq:geod-intermediate2}), which represents the corrections to the approximation in eq.~(\ref{eq:geod-approx}). Using $L \ll 1/\lambda$, which holds on the hyperbolic attractor, we obtain
\begin{align}
    \frac{\lambda}{L} \ll \frac{1}{L^2},
\end{align}
which yields
\begin{align}
    2L \log(\frac{\lambda}{6L}) < 2L \log(\frac{1}{6L^2}),
\end{align}
and as a consequence, using eq.~(\ref{ineq}), we obtain
\begin{align}
    \frac{2M_{pl}L \log(\frac{\lambda}{6L})}{\phi_f} &\lesssim \frac{2M_{pl}L\log(\frac{1}{6L^2})}{6M_{pl}(115)L}\nonumber\\
    &=\frac{\log(\frac{1}{6L^2})}{345} \label{eq:geod-correction}\\
    &\ll 1. \nonumber
\end{align}
for values of $L$ appropriate for our considerations. Note that while the expression in eq.~(\ref{eq:geod-correction}) does diverge in the small $L$ limit, this is beyond our concern since it is clear from eq.~(\ref{eq:geod-intermediate2}) that the absolute (as opposed to relative) size of the correction vanishes in the small $L$ limit. 

\section{Simplifying assumption}\label{app:simplifying}

Here we justify the assumption made in deriving the constraint in eq.~(\ref{eq:constraint}) on model parameters $\lambda$ and $L$, namely, that the two-field system $\varphi^I = (\phi,\psi)$ lies on the hyperbolic attractor solution described by eqs.~(\ref{eq:attractor1}) from today onwards.

\begin{figure}
    \centering
    \includegraphics[width=0.5\textwidth]{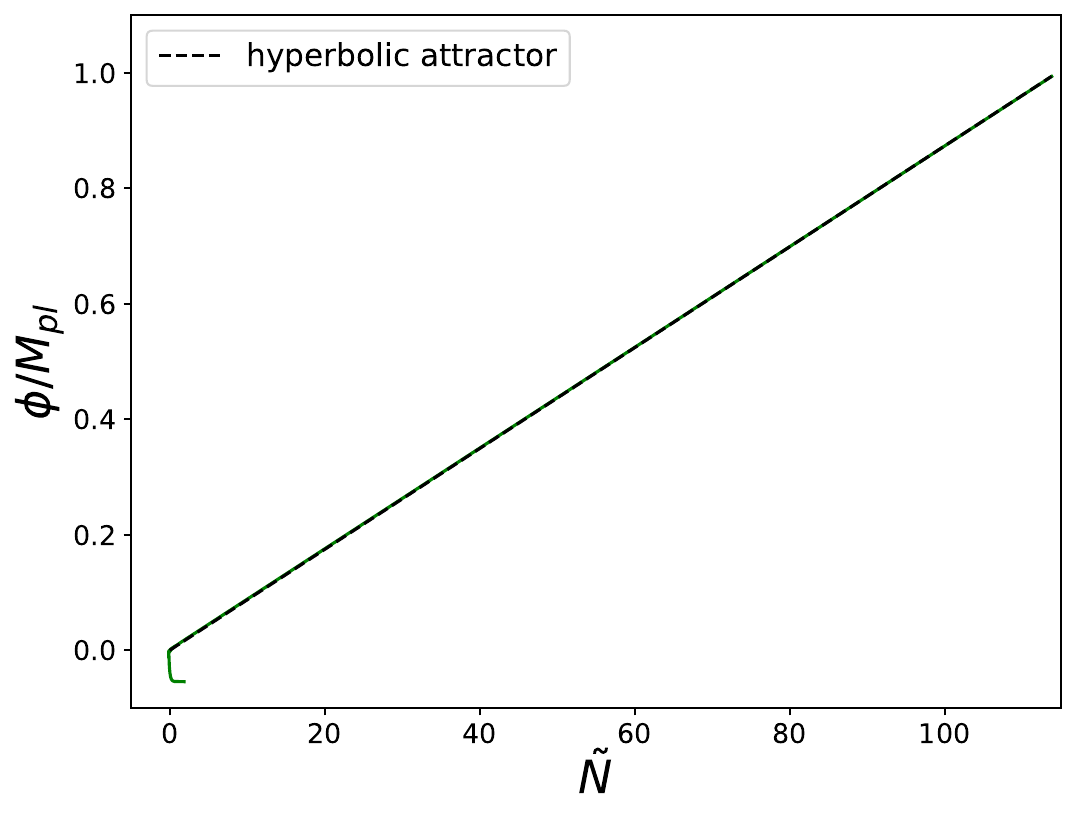}
    \caption{Transition of $\phi$ (green) from being frozen to reaching the hyperbolic attractor and evolution on the hyperbolic attractor. Assuming that $(\phi,\psi)$ evolves on the hyperbolic attractor from today onwards yields an excellent approximation (black).} 
    \label{fig:phi-transition}
\end{figure}

The basis for this assumption is that the transition period from when the fields $\phi$ and $\psi$ are frozen by Hubble friction to when they are evolving on the hyperbolic attractor lasts only a few e-folds, which is short in comparison to the number of e-folds $N \sim 115$ for which we consider the evolution of the system. Moreover, the field $\phi$ does not vary significantly in comparison to $M_{pl}$ during this transition. Figure \ref{fig:phi-transition} shows the evolution of $\phi$, obtained numerically, for $L=1/690$ and $\lambda = 1$, which demonstrates these two facts. This is a robust feature of this model in this cosmological setting. For more concreteness, we also plot in Figure \ref{fig:phi-transition} the approximate expression
\begin{align}
    \phi = \frac{12M_{pl}L}{-1-3w}\tilde{N},
\end{align}
(see eq.~(\ref{eq:phi-approx})) which comes from assuming that $(\phi,\psi)$ lies on the hyperbolic attractor from today onwards. The figure shows that the approximation is in excellent agreement with the numerics even at $\tilde{N} = 115$.

\end{document}